# Improved simulation of El Niño and its influence on the climate anomalies of the East Asia–western North Pacific in the ICM Version 2

HUANG Ping*[1,2], WANG Lei[1], WANG Pengfei[1,2], ZHANG Zhihua[3] and HUANG Gang[2,4]

[1]*Center for Monsoon System Research, Institute of Atmospheric Physics, Chinese Academy of Science, Beijing 100190*

[2]*State Key Laboratory of Numerical Modeling for Atmospheric Sciences and Geophysical Fluid Dynamics, Institute of Atmospheric Physics, Chinese Academy of Science, Beijing 100029*

[3]*National Marine Environmental Forecasting Center, Beijing 100081*

[4]*University of Chinese Academy of Sciences, Beijing 100049*

---

*Corresponding authors: HUANG Ping, huangping@mail.iap.ac.cn

# ABSTRACT

This study introduces the second version of the Integrated Climate Model (ICM). ICM is developed by the Center for Monsoon System Research, Institute of Atmospheric Physics to improve the short-term climate prediction of the East Asia–western North Pacific (EA–WNP). The main update of the second version of ICM (ICM.V2) relative to the first version (ICM.V1) is the improvement of the horizontal resolution of the atmospheric model from T31 spectral resolution (3.75° × 3.75°) to T63 (1.875° × 1.875°). As a result, some important factors for the short-term climate prediction of the EA–WNP is apparently improved from ICM.V1 to ICM.V2, including the climatological SST, the rainfall and circulation of the East Asian summer monsoon, and the variability and spatial pattern of ENSO. The impact of El Niño on the EA–WNP climate simulated in ICM.V2 is also improved with more realistic anticyclonic anomalies and precipitation pattern over the EA–WNP. The tropical Indian ocean capacitor effect and the WNP local air–sea interaction feedback, two popular mechanisms to explain the impact of El Niño on the EA–WNP climate is also realistically reproduced in ICM.V2, much improved relative to that in ICM.V1.

**Keywords:** Integrated Climate Model (ICM), El Niño, East Asian summer climate

# 1. Introduction

The seasonal prediction of the East Asian–western North Pacific (EA–WNP) climate is a long challenging task (e.g., Huang, 2006; Huang et al., 2007; Xie et al., 2016). Some well-known systematic biases of the simulation of EA–WNP climate have been reported in the state-of-the-art AOGCMs (e.g., Du et al., 2010; Brown et al., 2013; Sperber et al., 2013; Song and Zhou, 2014), hindering a more skillful seasonal prediction using numerical models. In 2014, the Center for Monsoon System Research, Institute of Atmospheric Physics (CMSR/IAP) released their first version of Integrated Climate Model (ICM) for improving the seasonal prediction of the EA–WNP climate (Huang et al., 2014; Wang, 2014). ICM is an atmosphere–ocean coupled general circulation model (AOGCM). In ICM, the atmospheric component is the Hamburg Atmospheric General Circulation Model Version 5 (ECHAM5) (Roeckner et al., 2003), the oceanic component is the Nucleus for European Modeling of the Ocean Version 2.3 (NEMO 2.3) (Madec, 2008) and the coupler is the Ocean Atmosphere Sea Ice Soil Version 3 (OASIS3) (Valcke, 2006).

The version 1 of ICM (ICM.V1) shows high skill to reproduce the climatology of sea surface temperature (SST) and precipitation in globe and the EA–WNP (Huang et al., 2014). The SST bias of ICM.V1 in the tropics is around 1°C–2°C, and ICM.V1 successfully reproduces the main characteristics of the seasonal cycle of SST in the equatorial Pacific—a famous challenge of AOGCMs. ICM.V1 also reproduce realistic strength and spatial pattern of the summer mei-yu rainband in the EA–WNP. The dominant interannual patten of the tropical Pacific simulated by ICM.V1 is the El Niño events associated with realistic pattern, amplitude and frequency. ICM.V1 can reproduce the East Asia–Pacific (EAP), or Pacific–Japan (PJ) pattern, which is the dominant teleconneciton pattern of the interannual anomalies of EA–WNP summer climate characterized by a meridional dipole with rainfall and circulation anomalies (Huang and Li, 1987; Nitta, 1987).

ICM.V1 can also reproduce a quite realistic process how the El Niño events influence the EA–WNP summer climate, which is one of the most important factors influencing the EA–WNP summer climate (Huang and Wu, 1989; Zhang et al., 1999; Wang et al., 2003; Huang et al., 2004; Xie et al., 2009; Huang and Huang, 2010; Xie et al., 2016). ICM.V12 can well reproduce the anomalous western Pacific anticyclone, the key anomaly system for the impact of El Niño on the EA–WNP climate anomalies. Two mechanisms popularly to explain the connection of western Pacific anticyclone and the El Niño, the local air–sea positive feedback (Wang et al., 2003) and the tropical Indian Ocean (TIO) capacitor effect (Xie et al., 2009), can also be realistically reproduced in ICM.V1. The local air–sea positive feedback by Wang et al. (2003) emphasizes the local positive feedback between the negative SSTAs and the low-level anticyclone in the WNP. On the other hand, the TIO capacitor effect by Xie et al. (2009) emphasizes the bridging role of the positive TIO SST anomalies following El Niño events. The positive TIO SST anomalies can continue through boreal summer and arouse anticyclonic anomalies in the WNP.

Because of the good performance of ICM.V1 in simulating the climate systems associated with the short-term climate prediction of the EA–WNP, ICM.V1 has been applied to the operational predictions in the CMSR/IAP and the National Marine Environmental

Forecasting Center of State Oceanic Administration of China. ICM.V1 has also successfully applied to investigate the role of the cloud–short-wave–SST feedback in the intermodel uncertainty of the relative change in the tropical Pacific SST under global warming (Ying and Huang, 2016).

However, some biases still appear in the simulation of ICM.V1, for example the excessive cold tongue and the double-ITCZ in the tropical Pacific, the excessively westward extended ENSO pattern, the overestimated climatological anticyclone over the WNP during boreal summer and the underestimated local air–sea positive feedback and the TIO capacitor effect (Huang et al., 2014). These biases could be associated with the low model resolution of ICM.V1, in which the horizontal resolution of the atmospheric component of ICM.V1 just is T31 spectral resolution (3.75° × 3.75°). Therefore, the ICM group of CMSR continues to develop the second version of ICM (ICM.V2). In ICM.V2, the horizontal resolution of the atmospheric component is upgraded from T31 to a higher T63, and some physical processes are improved. Because of these improvement, the performance of ICM.V2 has been further improved compared to ICM.V2 to reproduce the climatology and interannual variability of the EA–WNP climate. This study will report the improved performance of ICM.V2 in simulating the El Niño and its influence on the EA–WNP climate.

2. **Models, data and method**

ICM.V2 has the same framework with ICM.V1, which is briefly reviewed here. The details about the framework of ICM can be found in Huang et al. (2014) and Wang (2014). ICM consists of the atmospheric model ECHAM5 developed by the Max Planck Institute for Meteorology (Roeckner et al., 2003; Roeckner et al., 2006), the oceanic and sea ice model NEMO 2.3 by the Institut Pierre-Simon Laplace (IPSL) (Madec, 2008), and the coupler OASIS3 (Valcke, 2006). ECHAM5 transfers 17 variables including surface heat and freshwater flux and wind stress to NEMO, whereas NEMO transfers SST, three attributes of sea ice, and the speed of sea surface current to ECHAME5. The transferred variables are not changed from ICM.V1 to ICM.V2.

ICM.V1 uses the low-resolution version of ECHAM5 (T31 spectral resolution; horizontal 3.75° × 3.75° and vertical 19 levels). The horizontal resolution of ECHAM5 in ICM.V2 is upgrade to T63 spectral resolution (horizontal 1.875° × 1.875° and vertical 19 levels). The resolution of the oceanic part, NEMO, is not changed. The horizontal resolution of NEMO in ICM.V1 and V2 is around 2° at high latitudes and around 0.5° meridional resolution close to the equator, whereas the vertical resolution is 31 levels. The time step of the atmospheric model is improved from 2400 seconds in ICM.V1 to 1200 seconds in ICM.V2. The time step of the oceanic model is unchanged, 2400 second. The coupling frequency in ICM.V2 is the same as that in ICM.V1, once per 4 hours.

Some reanalysis datasets are analyzed to compare with the model simulation. They are the HadISST (Rayner et al., 2003), the NCEP/NCAR reanalysis 1 (Kalnay et al., 1996), and the Global Precipitation Climatology Project (GPCP) precipitation data (Adler et al., 2003). These datasets are called "observations" in the present study.

At present, ICM.V2 has steadily integrated 1000 years. A 300-year simulation in ICM.V2 is selected after a 700-year steady simulation (i.e. the period from the model year 701 to 1000) to compare with the 1981–2010 observations. Accordingly, the period from the model year 701–1000 in ICM.V1 is chosen for comparison. The time period of ICM.V1 is slightly different from the period in Huang et al. (2014), which induces slightly different results. The long-term mean of 300 years and 30 years define the climatologies in the modes and observations, respectively. To compare with the 30-year interannual variation in the observations, the total 300-year model outputs are divided into ten 30-year sections. The analysis is performed on each section, and then the averaged results of ten sections represent the results of the models.

Some results in ICM.V1 and observations are derived from Huang et al. (2014), which are shown in the present study to compare with the results in ICM.V2 conveniently.

3. **Climatology**

The seasonal mean SST during June–August (JJA) and December–February (DJF) simulated in ICM.V2 is shown in Figs. 1a and b. Compared with HadISST, the common bias of SST in 60°S–60°N is less than 1°C (Figs. 1c and d). The regions with bias around 1°C–2°C in ICM.V2 are much shrunken relative to those in ICM.V1 (Figs. 1e and f). Some typical biases of climatological SST in ICM.V1 are improved in ICM.V2, such as the excessive cold tongue, the warm biases on the flanks of the excessive cold tongue, the warm biases west the South Africa and America continents. All of them are common biases in many CGCMs (e.g. Mechoso et al., 1995; Ma et al., 1996; Davey et al., 2002; Gualdi et al., 2003; Luo et al., 2005; Park et al., 2009; Zheng et al., 2012; Li and Xie, 2014). The improved SST is corresponding to the improved surface wind, precipitation and coastal upwelling (not shown).

The seasonal cycle of the equatorial Pacific SST is important to the evolution of El Niño but cannot be realistically simulated in many models (e.g., Latif et al., 2001; Wang and Picaut, 2004; Guilyardi, 2006). Huang et al. (2014) shows ICM.V1 can quite well reproduce the seasonal cycle. Figure 2 shows the seasonal cycle of SST in equatorial Pacific deviating from the annual mean simulated in ICM.V1 and ICM.V2 and the observations. ICM.V2 can approximately reproduce the westward propagation of seasonal SST in the eastern Pacific, the annual cycle in the eastern Pacific and the semi-annual in the western Pacific. However, the simulation in ICM.V2 does not show apparent improvement relative to that in ICM.V1 (Figs. 2a and b).

The most important indicators of the simulation skill of one model for the EA–WNP summer monsoon are the skill reproducing the climatological distribution of precipitation and circulation (Du et al., 2010; Sperber et al., 2013; Song and Zhou, 2014). The EA–WNP summer climate simulated by ICM.V2 is shown in Fig. 3. The simulated mei-yu rainband in JJA shows realistic strength and spatial pattern (Figs. 3a and b). The overestimated WNP anticyclone and the associated northward extended mei-yu rainband in ICM.V1 (Fig. 3d) is improved in ICM.V2 (Fig. 3c). The overestimated precipitation over the Yangtze valley in ICM.V1 is also improved in ICM.V2.

4. **ENSO**

ENSO, the variation of tropical Pacific SST, is a dominant factor to the interannual variation of EA–WNP climate anomalies (e.g., Huang and Wu, 1989; Zhang et al., 1999; Wang and Picaut, 2004; Huang, 2006; Deser et al., 2010). Figure 4 shows the interannual variance simulated in the two versions of ICM and their biases compared with the observations. The main discrepancy in ICM.V1 and ICM.V2 relative to the observations is the modelled interannual variance extending too west, a common bias in the state-of-the-art models (Guilyardi et al., 2009; Collins et al., 2010; Christensen et al., 2013; Capotondi et al., 2015). However, the performance of ICM.V2 is much improved relative to ICM.V1 with smaller biases in the equatorial western and central Pacific (Figs. 4c and d).

Figure 5 shows the variance–period spectra of the Niño3 index in tropical Pacific SST. The regional averaged (5°S–5°N, 90°–150°W) SST anomalies (SSTAs) define the Niño3 index. The spectra of the Niño3 index in ICM.V2 are much closer to the observations than those in ICM.V1. The dominant variance of the Niño3 index in ICM.V2 is at a period of 3–4 years, which is closer to that in the observations (around 4 years) than that in ICM.V2 (around 3 years). The simulated interdecadal variation in ICM.V2 represented by the Niño3 index at the period longer than 4 years is much improved compared with that in ICM.V1, although it is yet weaker than the observations. This result indicates ICM.V2 has a higher skill in reproducing the interdecadal variability of the tropical Pacific SST.

The SSTA pattern of ENSO events is represented by the regression pattern of the tropical Pacific SSTAs onto the Niño3 index shown in Fig. 6. ICM.V2 can reproduce the main features of observed ENSO events with the maximum SSTAs in the equatorial central and eastern Pacific. The bias of the simulated SSTA pattern (Fig. 6c) displays as SSTAs extending farther west compared with the observations, which is a common bias in a lot of models (Collins et al., 2010; Vecchi and Wittenberg, 2010; Christensen et al., 2013), but this bias is much improved in ICM.V2 relative to that in ICM.V1 (Figs. 6c and d). The spatial distribution of the bias of ENSO SSTA pattern is similar to that of the interannual variance of SSTAs (Fig. 4). All the improved interannual variance, the spectra of the Niño3 index and the spatial pattern of ENSO SSTAs in ICM.V2 indicate that ICM.V2 performs much better than the first version to reproduce a realistic ENSO variability (e.g., Rasmusson and Carpenter, 1982). The improvement of ENSO variability in ICM.V2 could be associated with the improved climatological SST in the new version (Figs. 1 and 2).

5. **Impact of the El Niño events on the EA–WNP climate**

The skill of one model reproducing the influence of the El Niño events on the following summer climate anomalies over the EA–WNP is a key ability of the model to be utilized to the short-term climate prediction of the EA–WNP. The methods used in Huang et al. (2014) are also used in this study to evaluate the impact of ENSO on the following summer EA–WNP climate. The Niño3 index in the boreal winter mean [December–February; D(0)JF(1)] to represent the earlier signal of ENSO events. The anomalies of the EA–WNP rainfall and 850-hPa wind in the following July–August [JJA(1)] are regressed on the normalized D(0)JF(1) Niño3 index.

In the observations, the EA–WNP circulation anomalies induced by ENSO events (Fig. 7e) show pronounced anticyclonic anomalies over the WNP and cyclonic anomalies on the north of the anticyclone anomalies from northeast China to Japan. The EA–WNP rainfall anomalies also display a meridional dipole structure according with the circulation anomalies (Zhang et al., 1996; Zhang et al., 1999; Wang et al., 2003). In ICM.V1 (Fig. 7c), the structure of the anticyclone anomalies over the WNP is basically reproduced but with relatively smaller amplitude than the observations. The structure of the anticyclone anomalies is clearer in ICM.V2 (Fig. 7a) with much improved amplitude than in ICM.V1. Previous studies suggest that the WNP anticyclone anomalies following the El Niño events are associated with the positive SSTAs in the TIO and WNP, as illustrated by the boreal summer SSTAs regressed on the D(0)JF(1) Niño3 index in Fig. 7f (Zhang et al., 1996; Zhang et al., 1999; Wang et al., 2003; Xie et al., 2009; Xie et al., 2016). The positive SSTAs in the TIO and WNP is much weaker than the observations in ICM.V1 (Fig. 7d). In ICM.V2, the positive SSTAs in the TIO is much improved (Fig. 7b), whereas the positive SSTAs in the WNP are not improved much.

As introduced in the first section, there are two popular mechanisms to explain the influence of El Niño on the EA–WNP climate during the following boreal summer: the TIO capacitor effect (Du et al., 2009; Xie et al., 2009) and the local air–sea positive feedback (Wang et al., 2003). The observed TIO capacitor effect can be found from the 200-hPa potential height anomalies regressed on the D(0)JF(1) Niño3 index in Fig. 7e, exhibiting a Matsuno–Gill pattern over the warm pool with two Rossby-wave cycles over the western TIO and a Kelvin-wave trough over the tropical western Pacific (Du et al., 2009; Xie et al., 2009). Figure 8 shows the SSTAs and surface wind anomalies in the TIO from the boreal winter to the following summer regressed on the D(0)JF(1) Niño3 index, which are the key processes of the TIO capacitor effect (Xie et al., 2009). In the observations, the positive SSTAs appear in the south TIO originally induced by the atmospheric bridge effect related to El Niño and maintained by the downwelling oceanic Rossby wave. The interaction of surface wind anomalies and the climatological westerly winds can shift the maximum SSTAs from the south TIO to north and maintains the SSTAs to May (Xie et al., 2009).

As shown in Figs. 7a,b, the underestimated Matsuno–Gill pattern in ICM.V1 is much improved in ICM.V2 with a magnitude as large as that of the observations. ICM.V2 can reproduce the processes of the TIO capacitor effect much better than ICM.V1 (Figs. 8a,b). The improved aspects include the correlation of SSTAs with the Niño3 index, the strength of surface wind anomalies, and the shift of SSTAs from south to north. The underestimated post-ENSO TIO warming in ICM.V1 is a common bias in the state-of-the-art models, in which the TIO warming cannot continue through boreal summer as in the observations owing to some shortcomings in reproducing the air–sea interaction in the TIO (Du et al., 2013; Hu et al., 2014). The present evaluation implies that ICM.V2 has some advantages to describe the air–sea interaction in the TIO.

The local air–sea positive feedback (Wang et al., 2003) is also evaluated as shown in Fig. 9. The meridional-mean 850-hPa wind anomalies are regressed on the D(0)JF(1) Niño3 index, and the correlation between the SSTAs in the western Pacific and the D(0)JF(1) Niño3 index are also calculated. In the observations, the El Niño-induced wind anomalies exhibit an anomaly anticyclone over the WNP. Positive SSTAs appear under the northerly anomalies of the anomaly anticyclone, whereas negative SSTAs under the southerly anomalies from January(1) to July(1) (Fig. 9c). In ICM.V1, the anticyclone anomalies and SSTAs are not so strong as that in the observations, and the wind anomalies and SSTAs cannot persist through boreal summer. In ICM.V2, the strength of the SSTAs and wind anomalies is much improved, and the anomalies can persist through boreal summer as long as that in the observations. A new bias in ICM.V2 is the negative SSTAs appearing in 110°–150°E during June to August, which should be positive SSTAs in the observations.

## 6. Summary

The present study introduces the second version of ICM developed in CMSR/IAP/CAS. The atmospheric and oceanic components in ICM are ECHAM5 and NEMO 2.3, respectively, which are coupled by OASIS3 (Huang et al., 2014; Wang, 2014). The horizontal resolution of the atmospheric ECHAM5 in the first version of ICM is the T31 spectral resolution (horizontal resolution: 3.75° × 3.75°). The low resolution in ICM.V1 may be a key reason of some apparent biases in the first version. Therefore, we improve the horizontal resolution of ECHAM5 from the T31 to the T63 spectral resolution (horizontal resolution: 1.875° × 1.875°) as a new version of ICM, ICM.V2. The time step of the atmospheric model is also improved from 2400 seconds in ICM.V1 to 1200 seconds in ICM.V2 according to the improvement of the horizontal resolution. The new version of ICM has steadily run up to 1000 years. The present study evaluates the improvement in ICM.V2 relative to ICM.V1, including the global seasonal-mean SST, the seasonal cycle of the equatorial Pacific SST, the rain belt and circulation over the EA–WNP during boreal summer, ENSO, and the impact of ENSO on the following summer climate of the EA–WNP.

In ICM.V2, the climatological SST bias in 60°S–60°N is less than 1 °C compared with the observed climatology obtained from HadISST. This result is much improved relative to ICM.V1. Some well-known common bias in the state-of-the-art CGCMs and in ICM.V1, such as the double-ITCZ, the excessive cold tongue, and the warm biases along the east coast of South Africa and America are quite weak in ICM.V2. The climatological precipitation and wind during boreal summer in the EA–WNP is well reproduced in ICM.V2 with realistic strength and spatial pattern relative to the observations. The overestimated WNP anticyclone and the northward extended mei-yu rainband in the first version are also pronouncedly improved in ICM.V2.

The variability of the tropical Pacific SST has also been improved in ICM.V2 relative to ICM.V1. Compared with the simulations in ICM.V1, the dominant variance of the Niño3 index in ICM.V2 is at a period of 3–4 years, which is closer to that in the observations. Meanwhile, the underestimation of the interdecadal variation in ICM.V2 is improved. The bias of the farther west extending SSTAs associated with ENSO in ICM.V1—a common bias in a lot of models—is apparently improved in ICM.V2.

As introduced in Huang et al. (2014), ICM is mainly developed to improve the skill of short-term climate prediction of the EA–WNP. Huang et al. (2014) has shown that ICM.V1 shows quite high skill to reproduce how the El Niño events influence on the EA–WNP climate anomalies during the post summer (Zhang et al., 1996; Zhang et al., 1999; Wang et al., 2003; Du et al., 2009; Xie et al., 2009; Xie et al., 2016). Compared with the underestimated ENSO's impact on the EA–WNP climate to some degree in ICM.V1, the ICM.V2 can reproduce the impact of ENSO as strong as that in the observations, in which an abnormal anticyclonic circulation induced by El Niño is over the WNP and an abnormal cyclonic circulation on the north of the abnormal anticyclone. The improvement of the circulation anomalies induced by ENSO simulated in ICM.V2 could be associated with the better reproduced two mechanisms in ICM.V2 explaining the anomalies of the EA–WNP climate during the post-ENSO summer, the local air–sea interaction feedback (Wang et al., 2003) and the TIO capacitor effect (Xie et al., 2009). The simulated TIO warming in ICM.V2 can persist through boreal summer with realistic strength as in the observations. The ENSO-induced SSTAs and low-level circulation anomalies over the WNP is also more realistic in ICM.V2 than in ICM.V1.

The improvement of ENSO-induced TIO warming and WNP air–sea interaction feedback in ICM.V2 implies that ICM.V2 possess improved skill to reproduce the tropical air–sea interaction processes on interannual timescale. This skill of ICM.V2 is quite high compared with other CGCMs (Du et al., 2013; Hu et al., 2014; Wu and Zhou, 2015). This evaluation implies that the second version of ICM is a more reliable model than ICM.V1 to the short-term prediction of the EA–WNP summer climate.

***Acknowledgments.*** This study was supported by the Special Scientific Research Project for Public Welfare of State Oceanic Administration (Grant 201505013), the National Basic Research Program of China (Grant 2014CB953904), the National Natural Sciences Foundation of China (Grant 41425019, 41375112 and 41575088), the Youth Innovation Promotion Association of CAS and the Key Technology Talent Program of CAS. We thank the development groups of ECHAM5, NEMO and OASIS for providing the component models and coupler, which are used under their respective licenses.

**Figure Captions:**

Figure 1   Distribution of the simulated long-term mean SST in ICM.V2 in (a) JJA and (b) DJF. (c, d) The differences between the simulation in ICM.V2 and the observations (HadISST), and (e, f) the differences between the simulation in ICM.V1 and the observations.

Figure 2   Seasonal cycle of SST deviation from the annual mean over the equator. (a) The simulation in ICM.V2, (b) the simulation in ICM.V1 and (c) the observations.

Figure 3   Climatological precipitation and 850-hPa wind over the EA–WNP in JJA. (a) The simulation in ICM.V2, (b) the observations, (c) the difference between the simulation in ICM.V2 and the observations, and (d) the difference between the simulation in ICM.V1 and the observations.

Figure 4   Standard deviation of interannual anomalies of tropical Pacific SST in (a) the simulation in ICM.V2 and (b) the observations. (c) The differences between the simulation in ICM.V2 and the observations, and (d) the differences between the simulation in ICM.V1 and the observations. The results of the models and observations are based on the model-year period of 991–1000 and 1911–2010, respectively.

Figure 5   Variance-period spectra of the Niño3 index in the models and observations based on the model-year period of 991–1000 and 1911–2010, respectively.

Figure 6   Regression of tropical Pacific SSTAs onto the Niño3 index in (a) the simulation in ICM.V2 and (b) the observations. (c) The differences between the simulation in ICM.V2 and the observations, and (d) the differences between the simulation in ICM.V1 and the observations.

Figure 7   Correlation of JJA(1) precipitation anomalies (colors) and 200-hPa potential height anomalies (contours) with the D(0)JF(1) Niño3 index, and regression of 850-hPa wind anomalies (vectors) onto the D(0)JF(1) Niño3 index in the models and observations (left panels; a, ICM.V2; c, ICM.V1; e, observations). Correlation of JJA(1) SSTAs with the D(0)JF(1) Niño3 index in the models and observations (right panels; b, ICM.V2; d, ICM.V1; f, observations).

Figure 8   Correlation between the zonal-mean SSTAs in the tropical Indian Ocean (40°–100°E) and the D(0)JF(1) Niño3 index (contours), and regression of the zonal-mean 850-hPa wind anomalies over the tropical Indian Ocean on the D(0)JF(1) Niño3 index (vectors) in (a) the simulation in ICM.V2, (b) the simulation in ICM.V1 and (c) the observations.

Figure 9   Regression of the meridional-mean (5°–20°N) 850-hPa wind anomalies on the D(0)JF(1) Niño3 index (vectors) and correlation between the meridional-mean SSTAs and the D(0)JF(1) Niño3 index (contours) over the western North Pacific of (a) the simulation in ICM.V2, (b) the simulation in ICM.V1, and (c) the observations.

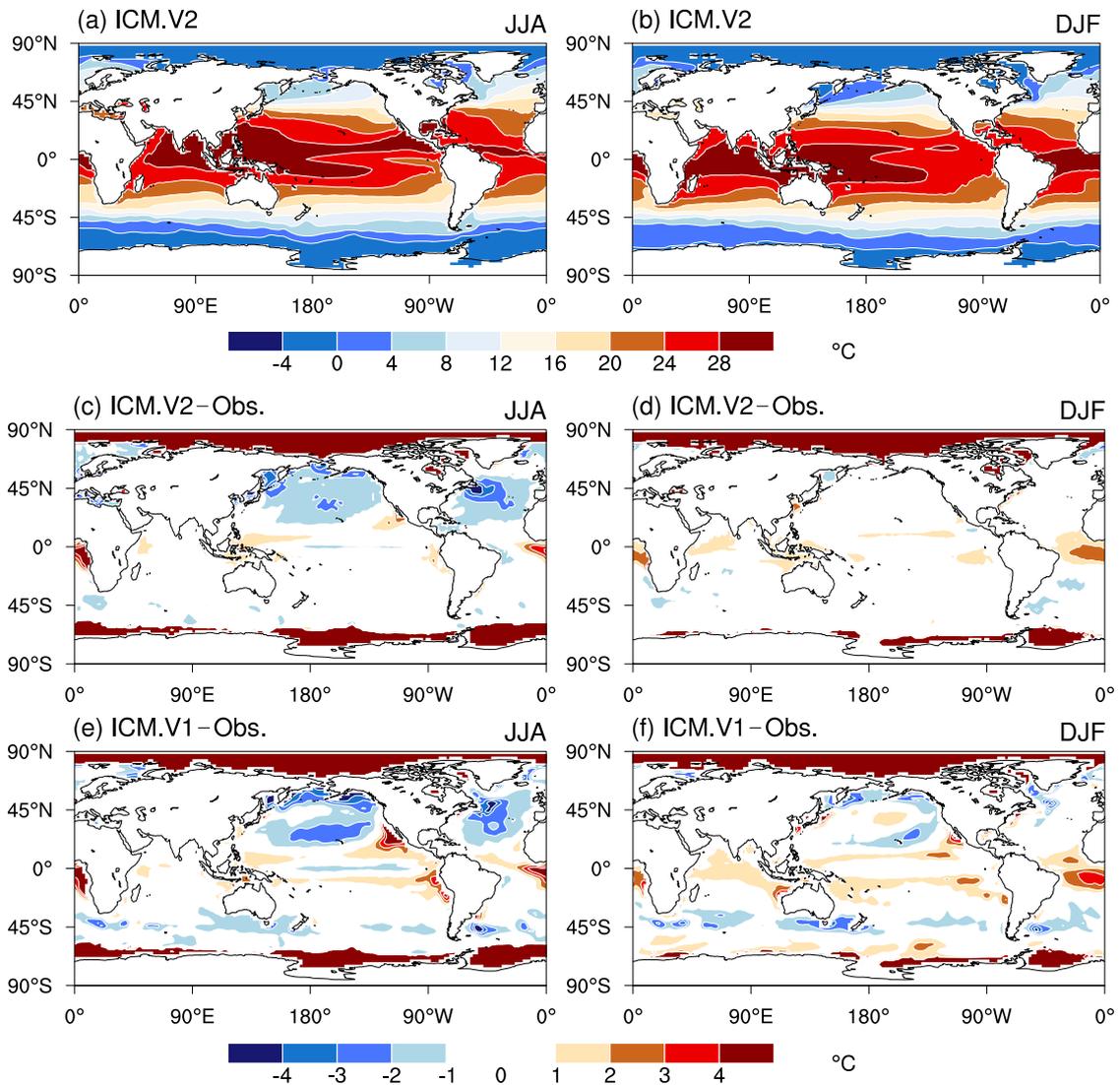

**Figure 1.** Distribution of the simulated long-term mean SST in ICM.V2 in (a) JJA and (b) DJF. (c, d) The differences between the simulation in ICM.V2 and the observations (HadISST), and (e, f) the differences between the simulation in ICM.V1 and the observations.

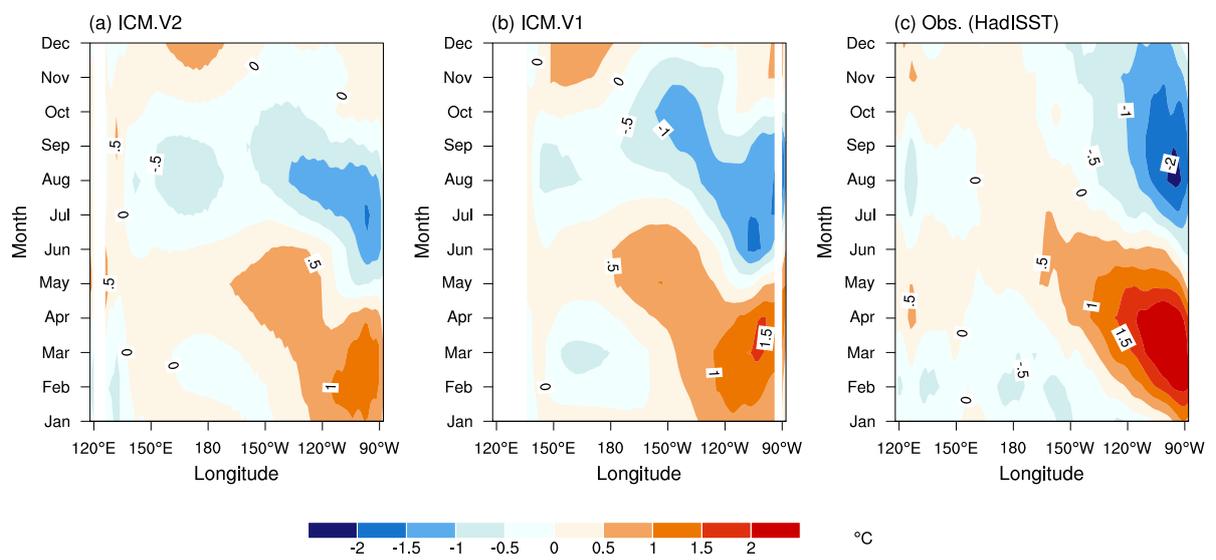

**Figure 2.** Seasonal cycle of SST deviation from the annual mean over the equator. (a) The simulation in ICM.V2, (b) the simulation in ICM.V1 and (c) the observations.

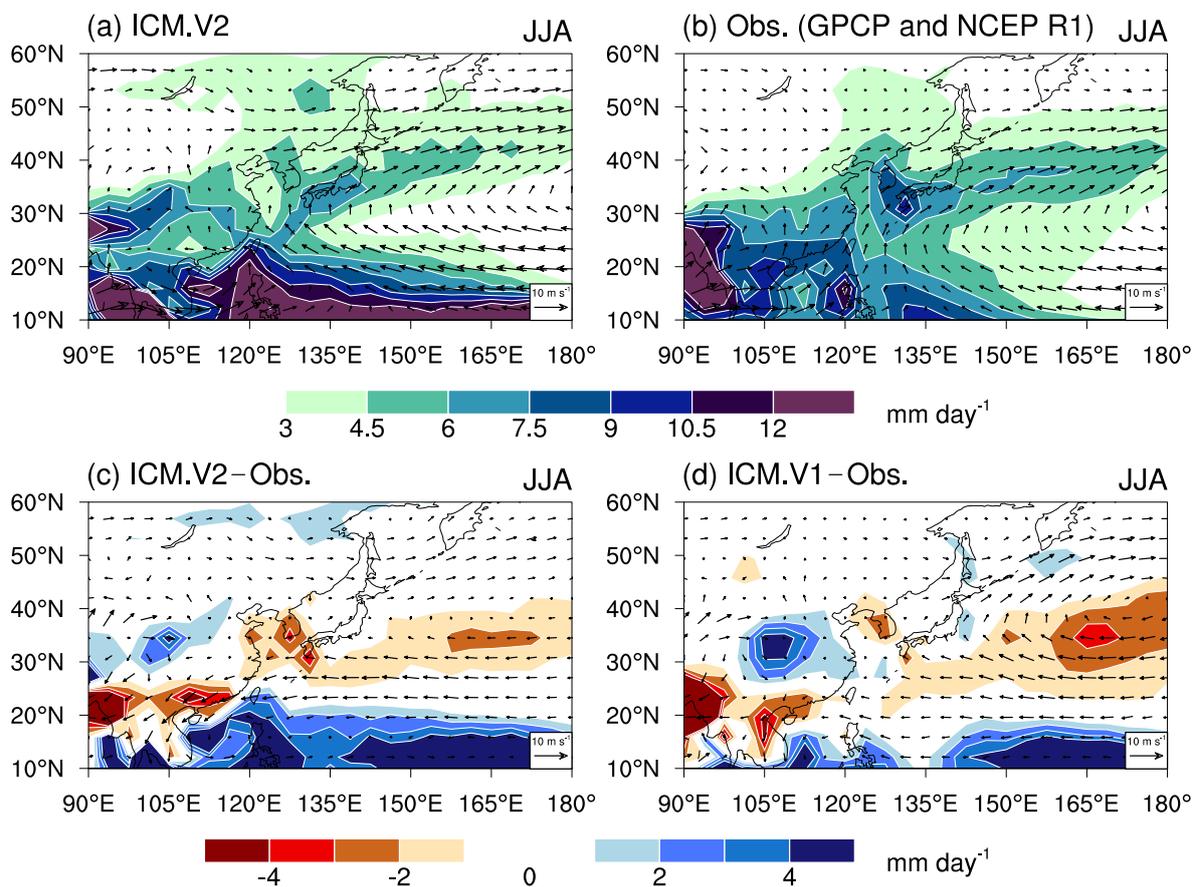

**Figure 3.** Climatological precipitation and 850-hPa wind over the EA–WNP in JJA. (a) The simulation in ICM.V2, (b) the observations, (c) the difference between the simulation in ICM.V2 and the observations, and (d) the difference between the simulation in ICM.V1 and the observations.

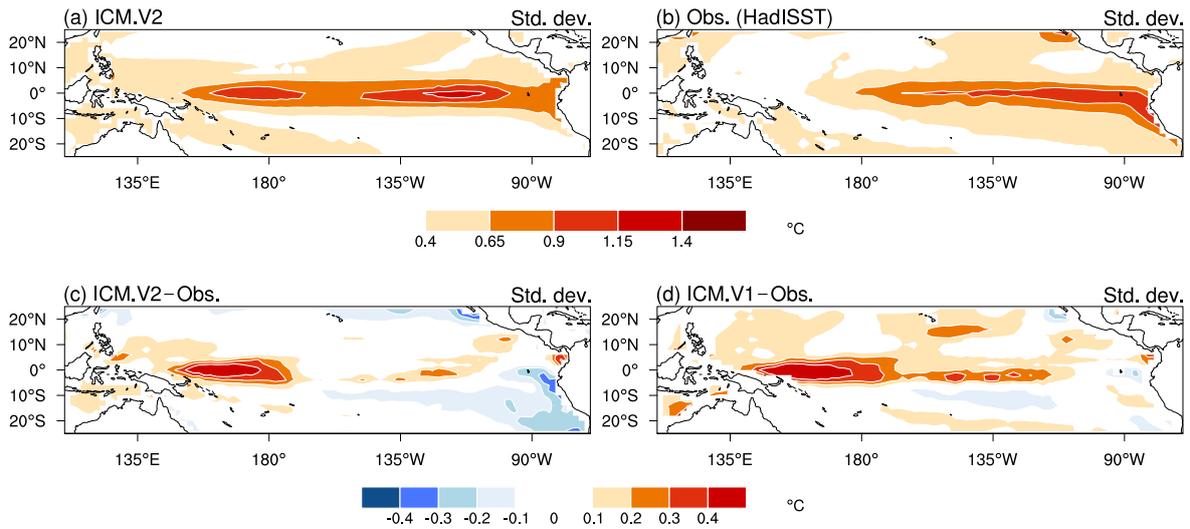

**Figure 4.** Standard deviation of interannual anomalies of tropical Pacific SST in (a) the simulation in ICM.V2 and (b) the observations. (c) The differences between the simulation in ICM.V2 and the observations, and (d) the differences between the simulation in ICM.V1 and the observations. The results of the models and observations are based on the model-year period of 991–1000 and 1911–2010, respectively.

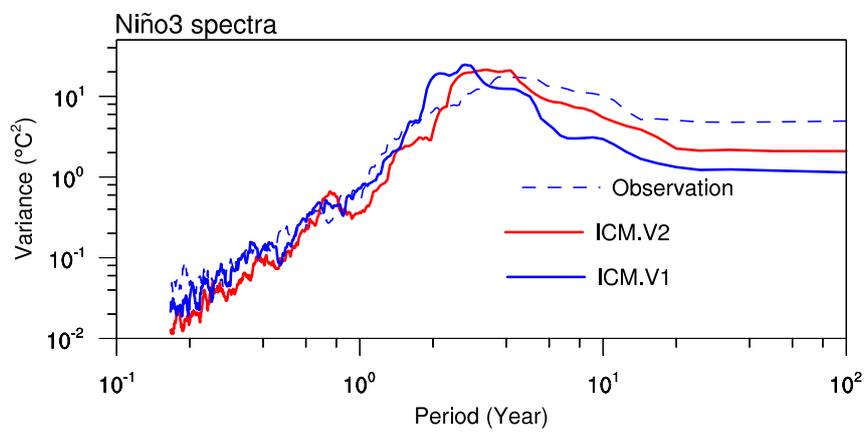

**Figure 5.** Variance-period spectra of the Niño3 index in the models and observations based on the model-year period of 991–1000 and 1911–2010, respectively.

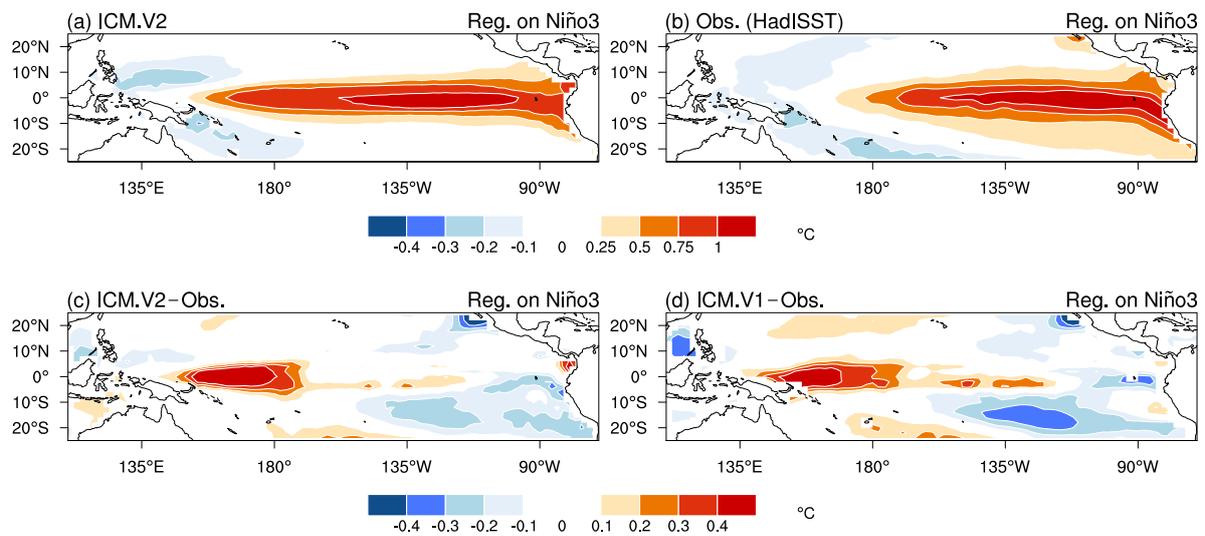

**Figure 6.** Regression of tropical Pacific SSTAs onto the Niño3 index in (a) the simulation in ICM.V2 and (b) the observations. (c) The differences between the simulation in ICM.V2 and the observations, and (d) the differences between the simulation in ICM.V1 and the observations.

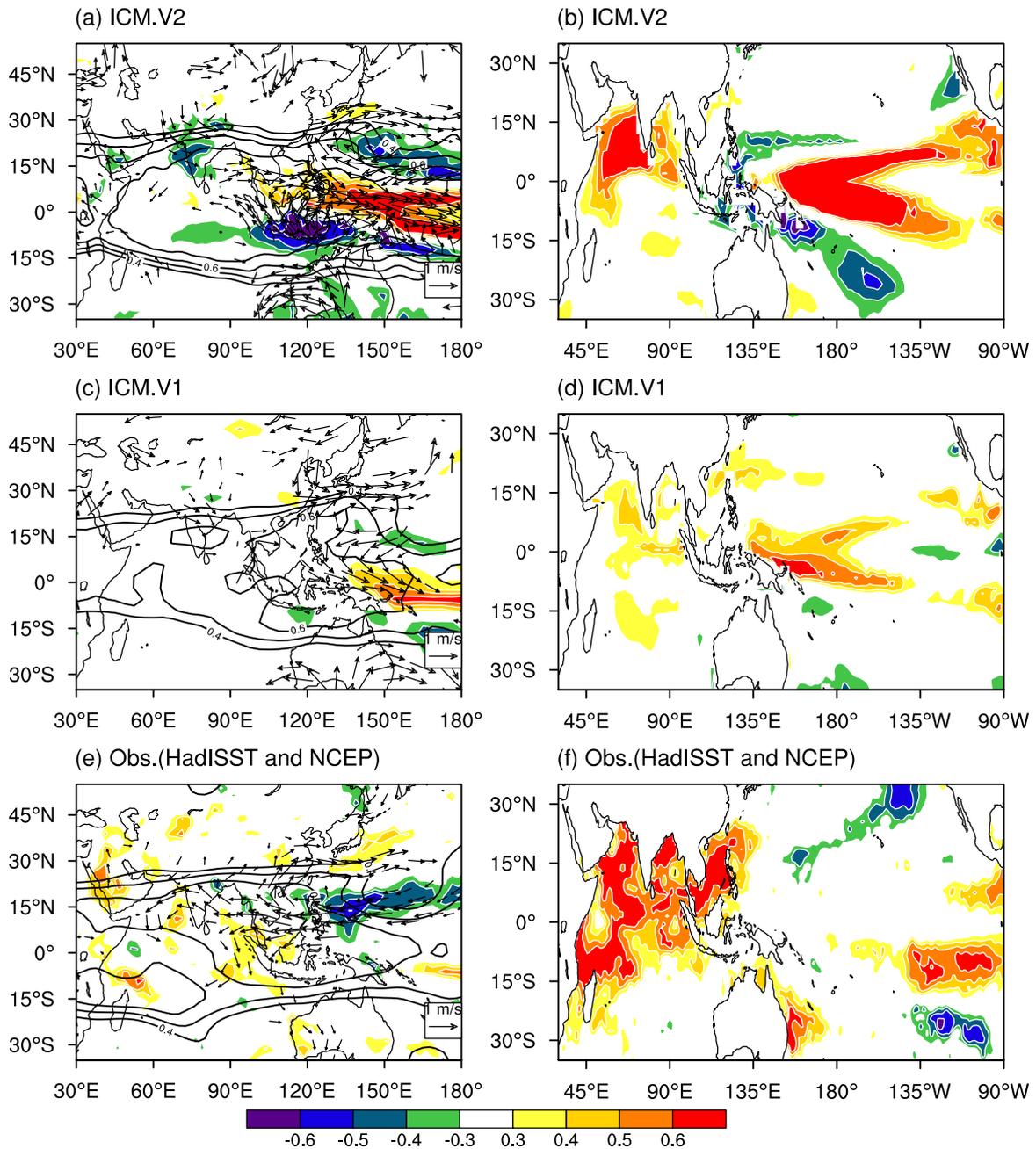

**Figure 7.** Correlation of JJA(1) precipitation anomalies (colors) and 200-hPa potential height anomalies (contours) with the D(0)JF(1) Niño3 index, and regression of 850-hPa wind anomalies (vectors) onto the D(0)JF(1) Niño3 index in the models and observations (left panels; a, ICM.V2; c, ICM.V1; e, observations). Correlation of JJA(1) SSTAs with the D(0)JF(1) Niño3 index in the models and observations (right panels; b, ICM.V2; d, ICM.V1; f, observations).

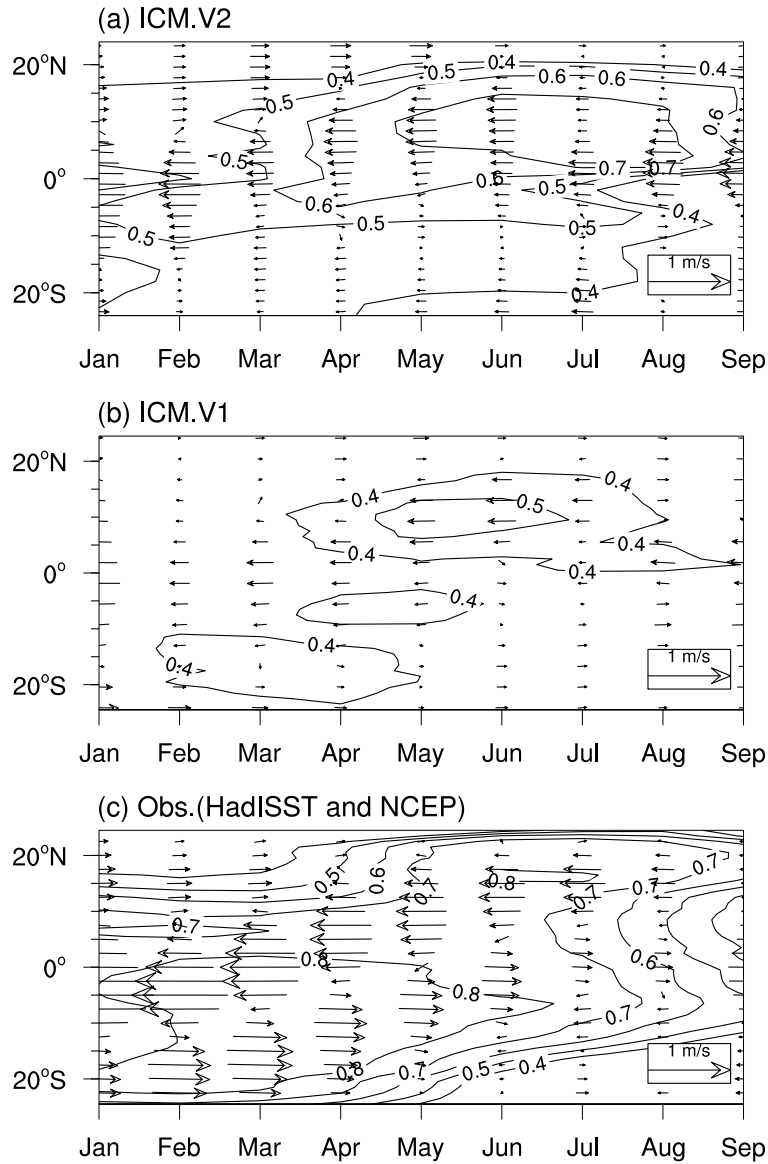

**Figure 8.** Correlation between the zonal-mean SSTAs in the tropical Indian Ocean (40°–100°E) and the D(0)JF(1) Niño3 index (contours), and regression of the zonal-mean 850-hPa wind anomalies over the tropical Indian Ocean on the D(0)JF(1) Niño3 index (vectors) in (a) the simulation in ICM.V2, (b) the simulation in ICM.V1 and (c) the observations.

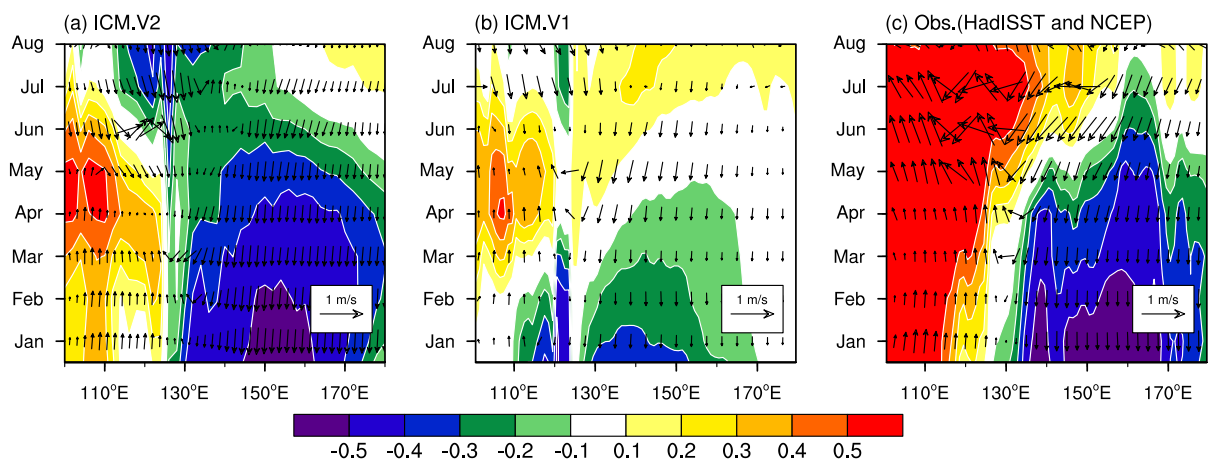

**Figure 9.** Regression of the meridional-mean (5°–20°N) 850-hPa wind anomalies on the D(0)JF(1) Niño3 index (vectors) and correlation between the meridional-mean SSTAs and the D(0)JF(1) Niño3 index (contours) over the western North Pacific of (a) the simulation in ICM.V2, (b) the simulation in ICM.V1, and (c) the observations.